\documentclass[submission,copyright,creativecommons]{eptcs}
\usepackage{iftex}
\ifpdf
  \usepackage{underscore}         
  \usepackage[T1]{fontenc}        
\else
  \usepackage{breakurl}           
\fi
\usepackage{paralist}   
\usepackage{graphicx}   

\title{Towards an Accessible\\Mathematics Working Environment\\
  Based on Isabelle/VSCode
}
\author{Klaus Miesenberger
  \institute{Johannes Kepler University \\ Linz Austria}
  \email{klaus.miesenberger@jku.at}
\and
  Walther Neuper
  \institute{Johannes Kepler University \\ Linz Austria}
  \email{walther.neuper@jku.at}
\and
  Bernhard St\"oger
  \institute{Johannes Kepler University \\ Linz Austria}
  \email{bernhard.stoeger@jku.at}
\and
  Makarius Wenzel
  \institute{Dr.~Wenzel \\ Augsburg Germany}
  \email{makarius@sketis.net}
}
\def\titlerunning{Accessible Mathematics Working Environment}
\def\authorrunning{Miesenberger, Neuper, St\"oger \& Wenzel}

\def\isac{\textit{ISAC}}
\def\sisac{\textit{ISAC}}

\begin{document}
\maketitle
\begin{abstract}
The paper collects preparatory work for interdisciplinary collaboration between three
partners, between (1) experts in improving accessibility of studies 
for impaired individuals, (2) experts in developing educational mathematics software
and (3) experts in designing and implementing interactive proof assistants.

The cooperation was started with the goal to develop an accessible mathematics
working environment for education with reasonable efforts. The start was
triggered by the lucky discovery that the upcoming Isabelle/VSCode is greatly
accessible for blind users with\emph{out} further impairments; this is envisaged
as the project's target group.

Technical details are described to an extent necessary to understand essential details
of efforts required for development. A survey of demand from
practice of education with respect to (1) and (2) leads to a vision for educational math
software, which necessarily is sketchy but suffices to guide development and
which shall invite experts in didactics of mathematics to collaborate.
\end{abstract}

\section{Introduction}
\emph{Isabelle} is a proof assistant \cite{Nipkow-Paulson-Wenzel:2002}
which is used worldwide by mathematicians to mechanise their theories
in order to make sure that they are correct. In that way much more
mathematical knowledge is mechanised than taught in undergraduate
courses at university.\footnote{\url{https://www.isa-afp.org/}}
\emph{VSCode}\footnote{\url{https://code.visualstudio.com/}} is a
source code editor for programmers made open source by
Microsoft. \emph{Isabelle/VSCode} is the proof assistant upcoming with
VSCode as a new front end instead of jEdit.

A happy coincidence led to the plan to build an accessible
mathematics working environment (in the sequel abbreviated MAWEN) within 
Isabelle/VSCode: The happy coincidence was that one of the authors found out 
that Isabelle/VSCode
is greatly \emph{accessible} for him, born blind and nevertheless an 
academic mathematician. First time in his life he could read proofs on a 
computer, by mouse-click on formulas he could look up theorems used in a proof 
and this simple way look up involved definitions as well\footnote{This person 
will play an important role in a planned project with his 
practical experience and theoretic expertise. Thus the project and this paper
will focus accessibility for gifted \emph{blind} persons only,
although the notion "accessible" also includes motor impairments and other kinds
of disabilities. In our estimation the restricted goal serves a major part of
possible target groups. So in this paper "accessible" is meant for blind
persons only.}.

Technical reasons for the idea to build a MAWEN based on Isabelle/VSCode
are straight forward since Isabelle has linear formulas: (special) characters 
like $\exists\; \forall \land \lor + - \;x_1$ or $x^2$
are fitted in one line in a sequence but $\frac{x+1}{y-2}$ 
is displayed as $(x+1)/(y-2)$ --- exactly the way
a visually impaired person encounters formulas on the Braille display!

\medskip
The happy coincidence was backed by long-term efforts at the Institute 
Integriert Studieren (IIS) of the Johannes Kepler University at 
Linz\footnote{\url{https://www.jku.at/en/institute-integriert-studieren/}}
with making mathematics studies accessible. Among other activities IIS cooperated
with the \sisac{} project\footnote{\url{https://isac.miraheze.org/}} 
which works on prototyping educational mathematical
software based on Isabelle for two decades.
Now it was straight forward to join with a third party, with Isabelle system 
development, and to start the MAWEN project. The joint project thoroughly 
investigated the demand for such software, detailed the user requirements and technical feasibility
and present all together in this paper. Technical feasibility has been checked at the Isabelle 
Workshop\footnote{\url{https://sketis.net/isabelle/isabelle-workshop-2022}},
details have been documented in \cite{IsaWS-22} and now
the project is underway to look for international cooperation with the
relevant academic institutions.

The project team already covers expertise in prover technology and accessibility as 
well as in practise of education and it looks forward to collaborate with
experts in educational science in the field. This will be important
since the design of the envisaged MAWEN is novel in one important aspect:
The actual MAWEN design claims that is covers \emph{all} aspects of doing
mathematical problem solving\footnote{Doing mathematics besides human 
interaction during learning,
also comprises geometry graphical proof methods filling multiple choice tests 
and the like; but this is already covered by other software products.} 
i.e. of modelling of problems of specifying 
knowledge required to solve a particular problem and last not least of 
step-wise solving the problem interactively with feedback from the system.
Since mathematics is \emph{the} science of reasoning where each statement and 
each step in a problem solution can and need to be justified,
reasoning should be implemented in the underlying software; this is done by 
the theorem prover \cite{EPTCS290.6} Isabelle, below shortened to ``prover''. 
This paper describes how
this is done already in the \sisac{} prototype and how further support for reasoning 
shall be implemented by the MAWEN project.

\medskip
In mathematics education better computer support appears particularly important 
for inquiry-based learning, because computers can make formulas interactively 
accessible (the respective potential in particular for inclusive learning will 
be shown in detail below) and learning formulas is the main entry point to the world of 
mathematics (already Euclid tried hard to grasp geometry by words, at that time
still lacking our elaborate language of mathematical formulas).

MAWEN's claim of covering \emph{all} activities in mathematical problem solving might be
questioned by experts in didactics of mathematics. In fact, the uncountably many
development efforts in software for mathematics education (also for disabled
persons) and the educational expertise developed around these software products
supersede the fact that mathematical problem solving is still done "by paper and pencil"
in schools today. Since the software most frequently used in school, GeoGebra,
is completely inaccessible and prominent mathematics software still presents formulas
as graphics (which is comprehensible only for sighted persons;
see for instance\footnote{\url{https://www.wolframalpha.com/}}) 
the situation for blind people
is particularly unsatisfactory. This paper will give directions to improve 
the situation.

\medskip
The paper is structured as follows: After the above introduction section 
\S\ref{access-math} clarifies the starting point for the MAWEN project
with respect to accessibility (\S\ref{state}), identifies challenges
to be met (\S\ref{challenges}) in order to derive specific user 
requirements (\S\ref{user-requ}). Section \S\ref{edu-soft-math}
addresses the many challenges for educational software in mathematics from three
sides: software limitations in flexibility and scope (\S\ref{tutor}),
the increasing role of (mechanical) proof in engineering (\S\ref{proof})
and justification of correctness in mathematical problem solving
(\S\ref{justify}). Section \S\ref{case-study} presents a case study 
investigating
how far mechanical support will be possible in solving complex mathematical
problems; the presentation distinguishes, which features have  tried and tested
already in the \sisac{} prototype and which are planned for the MAWEN project.
Section \S\ref{techno} turns to technology development in \S\ref{state-prover}, 
points at most recent software components envisaged for MAWEN in
 and in \S\ref{prospect-tech} and in \S\ref{open-technologies} asks 
questions to be tackled in the project.
Section \S\ref{edu-goal} takes the surveys of demand from \S\ref{access-math}
and from \S\ref{edu-soft-math} into account for a vision which drives MAWEN development.
Section \S\ref{conclusion} gives a conclusion and a summary.

\section{Accessibility in Mathematical Software}\label{access-math}
Major efforts have been undertaken to make the Internet accessible and the 
resulting browser technology, among others, is a great success. However 
mathematical content raises specific issues due to the structure of mathematical 
formulas, symbolic calculations and proofs. In computer mathematics the notion
of ``term'' is preferred over ``formula''; below, however, we use both notions 
depending on the context.

\subsection{Present State}\label{state}
The past four decades saw ever increasing benefit of blind and visually
impaired individuals from modern information technology. Unfortunately, however,
this benefit can not fully be extended to the area of mathematics. This
is mainly due to the following reasons:
\begin{compactitem}
\item Mathematical formulas are normally presented in a two-dimensional graphical
  manner which cannot be represented by adaptive tools for blind individuals
  which are confined to one-dimensional --- linear --- content, 
  see UR.1 on p.\pageref{UR-term} below.
\item Although there do exist quite a number of Braille codes for mathematics,
  which furnish a linear representation of math content, these codes are hard
  to learn, complicated, different between various countries and unknown to
  sighted people.
\item A platform to support collaborative work for blind and sighted persons
  in mathematics is missing.
\item There is only a very limited amount of mathematical literature available
  to blind individuals.
\item Blind people have tremendous difficulties when doing a mathematical
  derivation, because terms tend to become fairly complex at an early stage
  so that a blind person soon loses oversight over a symbolic calculation done by him
  or herself.
\item Until now there was no coherent software replacing ``paper and pencil'' 
  work in mathematical problem solving appropriately --- neither for blind nor for sighted people.\\
\end{compactitem}
	
There have been quite a lot of attempts to overcome these great
challenges; however, none of them is by now able to tackle the challenges
comprehensively and none of them was able to become widely used among pupils,
students and professionals who are concerned with mathematics and related areas.
Here is a brief list of examples for such attempts --- by no means complete but
somewhat representative:
	
\begin{compactitem}
\item The \emph{Infty Reader} \cite{inftyreader} is a mathematical
  OCR solution intended to remedy the above-mentioned lack of mathematical
  literature accessible to blind individuals. Although under the right circumstances
  the tool can produce quite accurate recognition results, the terms produced by
  it are extremely hard to read because they contain lots of formatting
  information which will distract the blind reader. Since OCR addresses
  urgent needs, the respective functionality is concern of ongoing research
  and deep learning techniques are outperforming the state-of-the-art
  \cite{kit-math-expr-OCR-2019}.
\item \emph{ChattyInfty} \cite{chattyinfty2,chattyinfty1} is a comprehensive mathematical
  editor which represents terms by synthetic speech. The problem is that
  listening to terms makes them hard to understand and a supplementation of the
  verbal representation by tactile output --- Braille --- is missing.
\item The Accessible Equation 
  Editor \small{\url{https://accessibility.pearson.com/resources/aee/index.php}}
  $\;$ furnishes mathematical editing capabilities
  with Braille support. The problem is that only a limited amount of
  mathematical Braille codes are supported which markedly limits the potential
  of the product to be spread over the world.
\item A PhD thesis carried out by Prajaks Jitngernamdan at the University of
  Linz Austria investigated the question whether assistants or wizards might
  support a pupil in doing elementary tasks in arithmetic 
  \cite{prajaks1,prajaks4,prajaks2,prajaks3,prajaks:phd}. By an assistant or
  wizard we mean a software program which guides the user through the various
  steps of a calculation giving him/her instructions and hints what to do in a
  particular step of a calculation. The calculations supported were confined to
  the four basic operations of arithmetic --- addition, subtraction,
  multiplication and division --- of floating-point numbers. The thesis showed
  that indeed a pupil could be supported by such a wizard, the problem being
  that the respective software granted too much support such that the pupil's
  ability to work on his/her own was not sufficiently trained.
\item A practical training work of Tony Bolkas also carried out at the
  University of Linz Austria implemented a software prototype which took a
  formula in \LaTeX, translated it into MathML content and pronounced it
  through a text-to-speech system (TTS) via synthetic speech. Special emphasis
  was laid on conveying the structure of a formula to the listener by inserting
  pauses at key points of the structural tree and by inserting special spoken
  mark-up keywords to signal the opening and closing of a structural element
  such as a fraction an index a root etc.\\
\par Our tests of the software prototype showed, that even if audio mark-up is  
present, mere listening to a term will not be sufficient for easily  
understanding it. What is needed in addition, is a mechanism to support  
traversing the term's structure; this is described in the next two
sections.
\end{compactitem}

\medskip
Several other promising attempts to overcome the huge problem of making the
computer usable for blind people to do mathematics have been done; but because
they were based on timed projects their development was discontinued when the
respective project was over.
	
\subsection{Open Challenges}\label{challenges}
As said, a well accessible and usable mathematical working environment for blind 
individuals is still lacking. As a consequence of the analysis given in the 
previous section, these are the most important open challenges to be overcome in 
order to furnish such a working environment:
\begin{compactitem}
\item To support the target group in understanding a complex mathematical term: 
Whereas sighted people are supported in understanding a term's structure by the 
traditional graphical mathematical presentation, the linear notations available to 
blind people do not facilitate quick overview over a complex structure. 
Therefore modern tools from software development such as collapse/expand 
outline view etc. have to be applied to linear mathematical notation in order to 
allow a blind user to quickly traverse a complex term. This should include the
possibility to easily decompose a term into its sub-terms to see the super-term 
of a term and to grasp the semantic meaning of an element of a term.

\item To design tools assisting a blind individual to do a mathematical 
calculation efficiently: Evidently, sighted people will organise their
calculations in a visual way which, however, cannot be directly carried over to 
blind users. Alternative methods to organise a calculation are required, 
especially to quickly jump to sub-calculations interim results etc; these 
methods have to be in accordance with the methods of structural analysis and 
traversal also referred to as ``navigation'' demanded above.
\item To facilitate collaborative work between blind and sighted individuals, 
typically sighted teacher and blind pupil on a mathematical task: If a blind 
individual were able to do mathematics efficiently by methods called for in the 
paragraphs above, but had no possibility to present his/her work to sighted 
persons such as teachers or school mates, we would have an incomplete solution. 
Conversely if a sighted person, typically a teacher, would not be able to 
easily show his/her mathematical exercises, explanations, etc. to a blind person, 
typically a pupil, \emph{inclusion} of blind individuals into mainstream learning 
could not take place.
\end{compactitem}

\subsection{User Requirements}\label{user-requ}
Here follows a set of user requirements stated earlier
\cite{nkarl:master,mahringer} to Isabelle/VSCode. User requirements
are prefixed by ``UR''.

\paragraph{UR.1: Decomposition of terms into proper sub-terms}\label{UR-term}
A sighted person takes benefit from the highly sophisticated notation for terms 
elaborated during centuries: exponents, subscripts, numerator and denominator 
above and below a horizontal fraction bar, symbols from various languages etc. 
All that can hardly be realised by a one-dimensional Braille display. However  
\cite{nkarl:master} showed by experiments that efficient navigation through 
sub-terms and super-terms \emph{can} be realised via keyboard and Braille 
display --- which supports understanding the structure of a term quite 
successfully. The (sub-)terms are represented as ASCII strings according to
specific Braille standards.

So there is the requirement that in parallel to the visual representation of 
terms (which looks nice in Isabelle's specific fonts) there is a proper term 
structure available for accessibility reasons --- and both representations 
should be in parallel for the purpose of inclusive learning.

\paragraph{UR.2: Semantic information for term elements:}\label{UR-semant}
Various support for requesting semantic information is well established in IDEs 
for a long time; in particular, getting a definition by clicking on a respective 
identifier in program code.  Isabelle/jEdit extends this feature to mathematical 
terms: A click, for instance on a plus operator in a term, gets the definition 
of a semi-group, the simplest algebraic structure with a plus operator.

This feature has been transferred from Isabelle/jEdit to Isabelle/VSCode --- and 
there it works perfectly accessible with keyboard shortcuts. High 
estimations of the very general tools provided by Isabelle also foster hopes 
that a click on a certain element on the screen might lead to different 
definitions and explanations depending on a user's level of mathematics 
knowledge.

\paragraph{UR.3: Survey on structures}\label{UR-struct}
of theories and of proofs is easily gained by a sighted person quickly scrolling 
up and down the screen --- while the Braille display is line oriented and each 
line has to be touched until a survey might become satisfying. 

For comprehension arbitrary switching between detail and survey is crucial. This 
requirement is well met by Isabelle/jEdit with a ``Sidekick'' parser and by 
nice support for collapsing and expanding respective branches in the tree 
structure of theories; a mouse click on a branch immediately displays the 
respective detail in a theory. Proofs in Isabelle/Isar \cite{Wenzel:2006:Festschrift} 
are visually structured by syntax highlighting and by indentation such that the 
structure is presented nicely by the overall visual impression.

Visually impaired persons need an approximation to all these useful features
for sighted users --- an approximation that replaces vision and mouse by
tactile sense, Braille and keyboard shortcuts.
One concrete minimal requirement is
an ``Outline view'' for theories and for proofs.

\section{Educational Software in Mathematics}\label{edu-soft-math}
There is an abundance of educational software available for mathematics,
also professional software is being used or at least explored in education, 
and still software for math education is produced all around the world.
Thus we have to ask: What is missing in available software? Does the still
ongoing production of education software for mathematics indicate that
some essence is missing there?

Since the planned project aims at inclusive learning, i.e. collaboration of
sighted and visually impaired students, this section starts from general considerations
on educational software in mathematics.

\subsection{Software and Mathematical Problem Solving}\label{tutor}
In Germany a much discussed public letter \cite{stellungn-dmv-gdm-mnu} on  
mathematics education complains that schools take apart mathematical problem solving
and split it up into different ``fundamental competencies'' \cite{PISA-math-2015}
such that too many students 
are unable to cope with complex mathematical problems encountered in studies for
engineering, biology, computer science, etc\footnote{Original in German: 
``Der Mathematikstoff wird nur h\"appchenweise 
angeboten und nicht ausreichend vernetzt: Aush\"ohlung, Entfachlichung, 
Entkernung des Mathematikunterrichtes sind das Resultat.''}. 
In many countries
similar complaints pushed educational institutions into efforts at the 
Secondary-Tertiary Transition (STT)\cite{EMS-STT-19}.

In analogy with splitting into separated competences in the theory of 
mathematics education, software tools used in mathematics education appear to 
split mathematical problem solving into separate parts as well: Computer Algebra Systems (also in 
hand-held devices) are used for equation solving, simplification and graphing; 
for calculations, statistics and graphing them spread-sheets are used. Dynamic-Geometry 
Systems are used for interactive construction and for checking true or false
geometry statements; myriads of applets demonstrate particular mathematical
facts etc.

On the other hand, mathematical problem solving generally is still done
``by paper and pencil'' in schools. We conclude from that fact that 
there are no tools general enough to cover the wide range of activities in
problem solving; the tools available are too limited in scope. Moreover the tools are
too limited in flexible interaction: \cite{prajaks:phd} demonstrated how 
difficult user guidance is even in simple numeric calculations. 
Projects aiming at such software tools, which cover a wide scope of problem
solving and also provide flexible interaction and user guidance (restrained in 
order to challenge students in trial \& error learning), e.g. Imps, 
\cite{farmer:eduIMPS} 
ActiveMath~\cite{ActiveMath-MAIN11}, eMath Studio~\cite{Ralph-Johan06a}, have 
not yet reached widespread use.

So the main challenges in mathematics education, i.e. to teach modelling 
problems in mathematical language and to solve those problems, appear not to be 
acceptably supported
by software, neither in school nor at the academic level. For the latter
level, however, there is at least one promising product~\cite{RISC5531}.

\medskip
Thus we identify a lack of appropriate software support in
mathematical problem solving, a lack which can be filled at the state of the art 
in computer mathematics, we claim. This paper serves to underpin the statement
by explaining the preparations for the MAWEN project.

\subsection{More Need for Proof in the Future}\label{proof}
Formal proof is the feature which distinguishes mathematics from other 
sciences,
thus the relevance of proof in mathematics education is well acknowledged 
\cite{ICMI19-2012}. 

But teaching to prove is considered a difficult part of mathematics education. 
At high-school teaching proof is limited, so strong efforts are 
required at the Secondary-Tertiary-Transition. There are many initiatives to 
support academic education in mathematical proof by software tools 
\cite{EPTCS328.1,EPTCS328.4,EPTCS313.3,EPTCS328.2,logic-tutors-11} and their 
number increases rapidly in the present, particularly in the field of computer 
science. Since the initiatives are driven by proactive lecturers, success is 
promising. 

The situation at engineering faculties is specific: there academic
education in general tries to bypass the difficulties of teaching to prove. 
So far this has not been considered a serious deficiency; however, since 
computer science more and more intrudes into other engineering disciplines, 
``Formal Methods'' \cite{DBLP:conf/fm/BjornerH14} become more and more important 
in order to cope with increasing complexity of technical systems by mathematical 
proof of crucial properties.

So an increasing demand in formal mathematics education is to be expected, which 
transgresses the borders of pure mathematics in academia. In the course of 
increasing the scope of mathematics education at engineering faculties, the 
above-mentioned difficulties with proving can be expected to pop up at a broad front. 

\medskip
Again we identify a lack of appropriate software support, this time of support
in early phases of teaching proof. One prerequisite of understanding proof
appears to be experience with matching: matching the premises of a theorem with a
concrete proof situation in order to decide, whether a theorem is applicable
or not. \S4.3 in \cite{IsaWS-22} gives an example of matching in early
mathematics education, in simple rewriting. Without familiarity with the 
mechanical nature, matching the notion of proof must remain somewhat esoteric,
in our opinion.

Teaching to prove is particularly difficult in early mathematics education,
since the interests and the mental maturity of pupils are different. 
It seems clear that teaching of proof cannot be 
done within some hours of lectures; rather, it takes time like learning a 
language, where mathematics is the most mechanical one among human languages.

Software support relieves the situation, because it allows to gain experience 
concurrently, even unconsciously while focusing other things in doing
mathematics --- if the software used implements the respective mechanisms of
reasoning. This again is planned by Isabelle/PIDE/MAWEN and we look forward
to evaluate the advantages we expect from this novel kind of software.

\medskip
Teaching mathematical proof to visually impaired students was not an issue so 
far, it was just out of scope of education in mathematics for people with special needs. 
Worldwide there is a handful of exceptional persons who were able to reach an 
academic degree in mathematics, last but not least struggling with a lack of 
tools for comprehending and manipulating mathematical terms.

\subsection{Justification and Transparency in Mathematics}\label{justify}
In the text above, there where two statements of lack of appropriate software for education
in mathematics --- now, where does our confidence come, that the gap can be 
actually filled, filled with investment of reasonable efforts? For visually
impaired students in particular? The confidence comes from the new
conceptual and technological base of interactive theorem proving and
respective mechanisation of mathematics.

\medskip
The \sisac{} project\footnote{\url{https://isac.miraheze.org/wiki/History}}
has been investigating possibilities to fill the gaps stated above for educational
math software for a long time; 
\cite{wn:proto-sys} gives a survey on the project's history from the technical 
side as well as from the user's side. After clarification over the years
\sisac's idea can be presented simply as follows: 

\begin{center}
\emph{Make software a completely transparent \& interactive model of 
mathematics\\ 
such that students can learn by trial \& error\\
all what is involved in solving complex problems in engineering.
}
\end{center}

\noindent
These are the explanations for the bold demand note:
\medskip
\begin{compactenum}
\item A software model of the whole content of a discipline is only possible 
  for the science of mathematics: mathematics can be an abstract game with 
  symbols, which can be implemented on computers completely and reliably.
  Note that this kind of model is \emph{not} a simulation as it would be for
  physical disciplines.
\item Learning by trial \& error requires user guidance which flexibly 
  restrains help, 
  such that a student is neither overwhelmed nor bored. Creating such
  user guidance automatically is a conceptual challenge. For the ``solve phase''
  this is accomplished by Lucas-Interpretation \cite{EPTCS-wn-20}
  \footnote{Peter Lucas was one of the pioneers in compiler construction
  \cite{pl:formal-lang-hist,pl:lucas70} 
  and Lucas-Interpretation (LI) was a late contribution to the field for education.
  In short words, LI interprets a program such that it proposes steps in a
  calculation \emph{and} relates user input to the calculation with the program.
  \url{https://de.wikipedia.org/wiki/Peter_Lucas_(Informatiker)}}
  For the ``specification phase'' such user guidance is
  described in the subsequent section~\S\ref{case-study}. This addresses the
  aspect of ``\textbf{interactive} model'' of problem solving.
\item\label{complete} The idea also claims for ``all what is involved in solving 
  problems''; this ``all'' involves a series of tasks addressing the aspect of 
  ``\textbf{complete} model'' in mathematical problem solving: 
  \begin{compactenum}
  \item\label{modelling} translating a situation from physical reality (in 
    textbooks described by figures and text) into mathematical formulas;
    this process is called the ``modelling phase'' in the sequel;
  \item\label{thy} identifying knowledge required to formulate a problem 
    mathematically:
    definitions, axioms and derived theorems --- to be found in ``theories'';
  \item identifying the type of a problem (called a ``problem pattern'' in the 
    sequel) at hand from a prepared collection, if the student wants to 
    benefit from user guidance;
  \item\label{LI} selecting a method able to solve the problem from a
    prepared collection containing the programs required for
    Lucas-Interpretation;
  \item\label{step} step-wise solving the problem similar to paper and
    pencil work, guided by Lucas-Inter\-pre\-ta\-tion; this is called the
    ``solve phase'' in the sequel;
  \item continuous switching between these tasks, for instance, during 
    Pt.\ref{modelling} looking up in accordance to Pt.\ref{thy} or from Pt.\ref{step}
    returning to Pt.\ref{modelling} for a sub-problem etc.
  \end{compactenum}
\item All knowledge handled in Pt.\ref{complete} is available in human readable
  format close to usual mathematical notation --- this is the aspect of 
  ``\textbf{transparent} model''.\\
\end{compactenum}
\noindent
The aspect of ``transparent model'' addresses also the claim for a gentle 
introduction of proof: Any time a student is interested in details during 
solving some engineering problem, the system reveals the underlying knowledge, 
for instance while doing a step in equation-solving in the ``solve phase'' 
Pt.\ref{step}, the student can ask the system which laws of algebra have been
applied in this step. Or the system knowing the next step according to 
Pt.\ref{LI} can ask the student to apply a certain law as suggested (automatically)
--- this is the task of the dialogue module for adaptive user guidance in the
\sisac{} prototype.

This kind of transparency also provides pre-stages of proof (by exposing
students gently to mechanical matching, see \S\ref{proof}), thus the headline 
announces ``justification'' instead of ``proof''. Such pre-stages also might pave
the way for introducing proof into engineering studies.

\section{Case Study: Solving Complex Problems}\label{case-study}
The case study illustrates the novel extent of (I) coverage of software support 
in learning mathematics implemented by the \sisac-prototype \cite{wn:proto-sys}.
The second novelty, the (II) technical features of Isabelle/VSCode supporting 
accessibility, is commented alongside.
(I) is not so obvious, 
not so obvious for mathematicians as for computer scientists who know about 
formal specification \cite{gries} as a prerequisite for automation by software.

The larger coverage of activities during learning (\emph{not} teaching)
mathematics is given by support for concrete problem solving. The 
\sisac{}-prototype was already able to demonstrate the larger coverage in
several field tests \cite{imst-htl07,imst-htl06,imst-hpts08}. A simple
reason prevents wide-spread application of \sisac: the availability of
formula-editors which support access to semantic information.
Now Isabelle/VSCode represents it's formulas such that they are a door to
the whole mathematics knowledge, the door to the semantics of the formal symbols
\cite{IsaWS-22}; this formula editor shall be adopted by \sisac. 
It should not be concealed, that Isabelle's formulas comprise
exponents, subscripts and virtually all mathematical characters
($\land, \Rightarrow, \sum, \int$, etc.) -- but they are a linear sequence
of characters on a line. This is, however, exactly what a blind person ``sees'' 
on a Braille-display; so the envisaged software will serve a distinctive target
group, blind students and their class-mates in inclusive learning.

With respect to (I) the case study will focus the initial part of problem 
solving, which mathematics teachers are used to treat less formally. In the 
field studies mentioned above the teachers skipped this part, because it was too
new for them and they did not want to expose their students to it. 
However, in education of engineers an important aim is to first clarify ``what'' 
the problem is (which usually cannot be captured completely by a \emph{formal} 
specification) and secondly consider ``how'' the problem to solve. Our case
study will focus the ``what'' in the so-called ``specification phase'', which 
in general is considered more difficult to cope with automation by software 
than the ``how'' in the so-called ``solution phase''. Our expectation
in the latter phase is documented in \cite{EPTCS-wn-20} and 
more generally in \cite{EPTCS290.6}.

The case study shall demonstrate, to what extent software support for learning
is feasible even in this crucial ``specification phase'', to what extent 
activities can be supported by software, activities which were up to 
``paper and pencil'' and thus were exclusive for blind students.
For instance, the Pt.\ref{modelling} in the previous section \S\ref{justify}
on ``modelling'' , is illustrated in \S\ref{formalise} below as part of the
specification. The large coverage leads to the hope
to overcome the separation into so many insular solutions with accessible
software, which do not help students in practice as complained in
\S\ref{access-math}.

\medskip
The presentation below uses the current implementation of \sisac's prototype.
In the presentation the extent of realisation is made explicit as well as ideas, 
which exceed the actual implementation and are up to future development\footnote{
Thus this case study concerns \emph{technical realisability} and 
\emph{not} a field test on usability.}.

\subsection{An Example from Electrical Engineering}\label{example}
The following text combined with a figure can be found in textbooks for electrical
engineering:\\

\begin{minipage}[c]{0.55\textwidth}
\textit{
The efficiency of an electrical coil depends on the mass of the kernel. 
Given is a kernel with cross-sectional area, determined by two rectangles of
the same shape as shown in the figure.\\
Given a radius $r=7$, determine the length~$u$ and width~$v$ of the rectangles
such that the cross-sectional area~$A$  (and thus the mass of the kernel) is a 
maximum.}\\
\end{minipage}
\hfill
\begin{minipage}{0.45\textwidth}
  \includegraphics[width=0.55\textwidth]{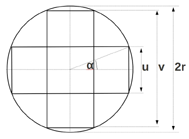}\\
\end{minipage}

\noindent
So what kind of problem is this? The word ``maximum'' indicates an optimisation
problem. Linear optimisation? What about approximation? Let us go deeper into 
details.

\subsection{From Text and Figure to Formulas}\label{formalise}
The text asks for a ``maximum''. In order to ``Find'' a maximum, what do we know
already, which facts are ``Given''? The following template has fields to enter
respective items in lines \rm{05} and \rm{07} respectively (where the line 
numbers serve for reference only on paper here):\label{expl}

{\footnotesize \begin{tabbing}
12345\=\rm{00}12\=12\=12\=12\=1234\=12\kill
\>\rm{01}\>Example $<$No.123 a$>$              \\
\>\rm{02}\>Problem ``univariate\_calculus/Optimisation''     \\
\>\rm{03}\>\>Specification:                 \\
\>\rm{04}\>\>\>Model:                      \\
\>\rm{05}\>\>\>\>Given: ``Constants'' $\;$ \framebox{$r=7$}\\
\>\rm{06}\>\>\>\>Where: $0 < r$ \\
\>\rm{07}\>\>\>\>Find: ``Maximum'' $\;$\framebox{$A$} ``AdditionalValues'' $\;$\framebox{$u v$}\\
\>\rm{08}\>\>\>\>Relate: \framebox{$A=2\cdot u\cdot v-u^2\;(\frac{u}{2})^2+(\frac{v}{2})^2=r^2$} \\
\>\rm{09}\>\>\>References:                 \\
\>\rm{10}\>\>\>\> \>RTheory: \framebox{``Diff\_App''}      \\
\>\rm{11}\>\>\>\>\framebox{x}\>RProblem: \framebox{``univariate\_calculus/Optimisation''}\\
\>\rm{12}\>\>\>\>\framebox{o}\>RMethod: \framebox{``Optimisation/by\_univariate\_calculus''}\\
\>\rm{13}\>\>Solution:
\end{tabbing}}

\noindent
The above template is already filled completely (frame-boxes indicate inputs) in order to 
shortcut explanation to the experienced reader. In the \sisac{} prototype the
template is offered to students at the beginning after she has chosen a
problem, for instance the one in \S\ref{example}. The \emph{``Specification''}
comprises a ``Model'' (line \rm{04}) and ``References''
(line \rm{09}), the former accepting input of what is found immediately in the
problem statement and the latter referencing knowledge already present in the
system, which could be helpful in solving the problem. Inputs to ``Model''
and ``References'' can be done in parallel.

 There are three kinds of knowledge (all with the leading letter 
``R'' for technical reasons and check-boxes ``x'' and ``o'' explained later): 
``RTheory'' collects all the types and definitions required to get the formulas 
in the ``Model'' accepted by the system. 
``RProblem'' is the identifier of the problem-pattern generating the
specific problem at hand. 
``RMethod'' is able to generate the ``Solution'' (line \rm{13}) 
automatically by Lucas-Interpretation
or in interaction with the student --- which is collapsed and not
discussed here as noted above.

The inputs on the specification template above depend on each other: For instance if the
student decides for an ``RTheory'', where the notion of fraction is unknown,
some formulas in line \rm{08} would report a syntax error.

\paragraph{The post-condition} requires specific consideration, in particular 
for readers experienced in formal methods: A problem is most accurately 
characterised by its post-condition. In case of optimisation problems a 
post-condition is like this:
{\footnotesize \begin{tabbing}
12345678\=$\forall \;A^\prime\; u^\prime \;v^\prime.\;$
       \=$\;(A^\prime=2\cdot u^\prime\cdot  v^\prime-(u^\prime)^2$\=123\kill
\>\>$\;A=2\cdot u\cdot v-u^2 \;$
    \>$\land\;\; (\frac{u}{2})^2+(\frac{v}{2})^2=r^2 \;\;\;\;\;\land$\\
\>$\forall \;A^\prime\; u^\prime \;v^\prime.\;(A^\prime=2\cdot u^\prime\cdot  v^\prime-(u^\prime)^2$
    \>\>$\land\;\; (\frac{u^\prime}{2})^2+(\frac{v^\prime}{2})^2=r^2) \Longrightarrow A^\prime \leq A$
\end{tabbing}}
A lecturer experienced in teaching students at engineering faculties knows, that
students would be generally overwhelmed by such formal statements, although they most 
precisely record insights gained from the figure: For the area $A$ of the 
cross-shaped kernel we have $A=2\cdot u\cdot v-u^2$, i.e. the area of a 
rectangle $u\cdot v$ times two, minus the overlay of the second rectangle. 
The second part $(\frac{u}{2})^2+(\frac{v}{2})^2=r^2$ contains already a 
choice of the student, which will be discussed in the sequel.

Due to the complexity of many post-conditions, the \sisac{} design decided to
encounter students with the operative minimum of the post-condition as shown by
``Relate'' in line~\rm{08}.

\paragraph{Extent of realisation and accessibility} of the post-condition in the prototype is
minimal. The proof of the post-condition must be done by automated provers,
which are present in Isabelle \cite{blanchette:sledgehammer} but not yet
employed in \sisac. A particular challenge is a pattern language, which
allows to instantiate a post-condition, formulated in patterns general for a whole
problem class, with a formalisation.

Accessibility design will be concerned with features for easily jumping through
the various fields of a specification; but the formulas in this phase are still
simple.

\subsection{``Next-Step Guidance'' by a Hidden Formalisation}\label{formalisation}
As already indicated just above, the general insight about right angles can be 
formalised by the sides' lengths in the Pythagorean triangle as 
$(\frac{u}{2})^2+(\frac{v}{2})^2=r^2$, but this can be done at least by one other
equivalent formula $(\sin\alpha)^2 + (\cos\alpha)^2 = r^2$. A student should
be supported in all possible ideas, how to solve the problem at hand --- by 
``Next-Step Guidance''~\cite{gdaroczy-EP-13,wn:lucas-interp-12}: the system 
should always be able to propose a next step to the student, who is free to 
adopt it or to neglect it.

Modelling the problem at hand we already have at least two 
choices for the student --- how to support these by the system? For that purpose
\sisac's prototype introduces ``hidden formalisations'' as follows (where the
``References'' are omitted, which are the same in both cases here):
{\footnotesize \begin{tabbing}
12345678\=$\forall \;A^\prime\; u^\prime \;v^\prime.\;$
       \=$\;(A^\prime=2u^\prime v^\prime-(u^\prime)^2$\=123\kill
\>$F_I\;\;\equiv\;\;[\; [r=7]  \;[A \;[u\;v]]\; 
   [A=2\cdot u\cdot v-u^2\;(\frac{u}{2})^2+(\frac{v}{2})^2=r^2]\;\{0<..<r\}\;]$\\
\>$F_{II}\;\;\equiv\;\;[\; [r=7]  \;[A \;\alpha]\; 
   [A=2\cdot u\cdot v-u^2\;\frac{u}{2}=r\sin\alpha\frac{v}{2}=r\cos\alpha]\;\{0<..<\frac{\pi}{2}\}\;]$\\
\end{tabbing}}

\noindent
That knowledge\footnote{The formalisations use brackets ``['' and ``]'' 
denoting lists as usual in functional programming. Some inner lists actually 
are interpreted as sets.}
is hidden behind the example (p.\pageref{expl} line \rm{01})
as for each example, which should be capable of ``Next-Step Guidance'' ---
an additional authoring effort, small given the fact, that user guidance in
interactive problem solving is generated automatically afterwards.

\paragraph{Modelling} is the process of translating ideas somehow 
represented in the mind or by some problem statement in an example
into a formal representation. How to guide that highly creative and intuitive 
process, that is now evident given the hidden formalisation: The system has simply to
match user input with the respective items in the formalisations. This allows
for reliable feedback; an item input to the example's ``Model'' can get these 
kinds of feedback:
\begin{compactitem}
\item \textit{Correct}
\item \textit{Superfluous}, for instance $u^2 + v^2 = (2r)^2$ in addition to
  $(\frac{u}{2})^2+(\frac{v}{2})^2=r^2$
\item \textit{Missing}, for instance ``AdditionalValues'', because these are
  required for the specific ``RMethod: Optimisation/by\_univariate calculus''  
\item \textit{Incomplete}, as long as one of the items is missing
\item \textit{SyntaxError}
\item \textit{False}, only for the pre-conditions in ``Where'' as long as
  $r=7$ is not input.
\end{compactitem}
\medskip\noindent
Matching user input also allows to rename identifiers, if wanted by students.

\paragraph{Guards for the methods} are of the same structure as 
``Model'' for problems. For these formalisations also may contain additional
items; for instance, note the two different items $\{0<..<r\}$ and 
$\{0<..<\frac{\pi}{2}\}$ in $F_I$ and $F_{II}$ respectively: These open 
intervals determine the range, where the solutions of equation solving are 
meaningful with respect to physical conditions; the calculation 
on p.\pageref{calc} below shows the use of this interval.

At this point we can also explain the check-boxes ``x'' and ``o'' preceding 
``RProblem'' and ``RMethod'': On p.\pageref{expl} on line \rm{11} the former is checked by ``x''
and the latter is not; that means that the ``Model'' shows items for 
``RProblem''. Otherwise checking ``RMethod'' would reveal the additionally 
``Given'' interval either $\{0<..<r\}$ or $\{0<..<\frac{\pi}{2}\}$
depending on what kind of ``Model'' is actually preferred by the student's
input.

\medskip
So the envisaged system has to be able to cope with a list of formalisations 
for one example under construction and to select the appropriate one, which the 
students seems to have in mind actually (which can change by specific new input).

%

\subsection{Relate the Problem to Available Knowledge}\label{references}
The three kinds of knowledge mechanised in the \sisac{} prototype have already
been mentioned; now they are introduced properly --- and related to 
accessibility, which becomes relevant first time in this case study.
\begin{enumerate}
\item\label{theories}\textit{Theories} are inherited from Isabelle without change. 
  Here are all the definitions required to mechanically read formulas 
  and all theorems required to evaluate the pre-conditions in ``Where''
  and to justify the steps in ``Solution'' after the ``Specification''-phase has been 
  completed. Theories are collected in directed acyclic graphs (DAGs) --- 
  there are general tools for displaying and manipulating DAGs, and here
  is the first specific challenge for implementing accessibility: To the best
  knowledge there is no \emph{accessible} tool available.
  This lack of appropriate tools will cause considerable efforts in 
  user requirements capture as well as in design and implementation.
\item\label{problems}\textit{Problems} are instantiations of a problem-pattern 
  with a formalisation. 
  Their representation has been introduced on p.\pageref{expl}.
  Their definition is given in theories as soon as these comprise the required
  knowledge. The structure of a problem collection
  is a tree --- and there are lots of tools 
  with highly elaborated accessibility, triggered by implementations of 
  files-browsers in any operating system.
\item\label{methods}\textit{Methods} comprise a guard, as mentioned at the end of
  \S\ref{formalisation}, and a program, which supports interactive construction
  of a solution for the specified problem~\cite{EPTCS-wn-20}.
   While an optimal structure for collections of methods is still unclear,
  the \sisac{} prototype collects them in a tree the same way as problems. 
  So the notes on accessibility apply here as well.
\end{enumerate}

\noindent
Searching appropriate knowledge during problem solving is probably the task,
which takes most advantage from software support. Most of the various tools
for such search are highly accessible as mentioned above. One exception is the
directed acyclic graph of theories; but any advance gained for accessible
display and manipulation of DAGs will bring benefits for the 
whole Isabelle community.

\begin{figure} [htb]
  \centering
  \includegraphics[width=0.75\textwidth]{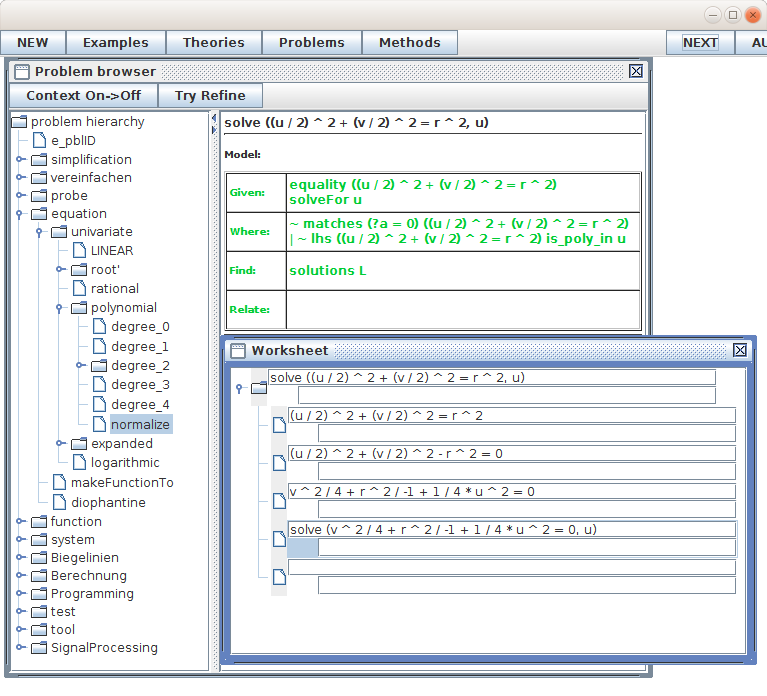}
  \caption{Tree of problems helpful during equation solving.}
  \label{fig:problem-tree}
\end{figure}

Fig.\ref{fig:problem-tree} on p.\pageref{fig:problem-tree}shows how searching the problem tree helped
during equation solving in the \sisac{} prototype with the old interface to be
replaced by Isabelle/VSCode \cite{Wenzel:2019:MKM}. The equation is a preview 
to p.\pageref{calc} referring to the step from line \rm{06} to line \rm{07}.
The snapshot is taken in a situation, where the student already succeeded in
normalisation of the equation (transformation to $\it{lhs}=0$); now the question is,
which type of equation to select (together with the method to proceed) ---
clicking an item in the tree triggers a check of the respective pre-conditions in ``Where''
and gives feedback by red colouring of the preconditions evaluating to false.

\paragraph{Extent of realisation and accessibility:} Extent of the knowledge 
Pt.\ref{theories}..Pt.\ref{methods} available in the prototype was sufficient 
for the field tests \cite{imst-htl06}, \cite{imst-htl07} and  \cite{imst-hpts08};
presently the knowledge comprises simplification of rational terms,
elementary differentiation, integration and equation solving on fractions and
square roots. For use of MAWEN in educational practice it will be necessary, 
that the knowledge covers major parts of the curriculum for the classes it is 
introduced. While this case
study shows usefulness for engineering studies, knowledge will be implemented
ascending with the curriculum from seventh grade of high school on, the time of 
introducing variables in mathematics education.

Authoring tools are still weak in the prototype; respective development will
gain major attention in the MAWEN project, since specific knowledge is a
prerequisite for next-step guidance.
Accessibility seems no challenge here, since the presentation of knowledge
will employ VSCode's standard tools, and these are perfectly accessible.

\subsection{Iterating through the Tasks}\label{iterating}
Problem solving is an iterative process of trial and error
between search and application of knowledge in the ``Specification'' phase and
between experiment and success in the subsequent ``Solution'' phase.
This iterative nature is reflected in the software-supported process as well ---
see a part of the ``Solution'' below following the completion of the
``Specification'' which is $<$\textit{collapsed}$>$ already):\label{calc}

{\footnotesize
\begin{tabbing}
1234\=\rm{00}12\=123\=123\=123\=123\=123\=123\=\kill
\>\rm{01}\>Example $<$No.123 a$>$                        \\
\>\rm{02}\>Problem ``univariate\_calculus/Optimisation'' \\
\>\rm{03}\>\>Specification:       \`\it{$<$collapsed$>$} \\
\>\rm{04}\>\>Solution:                                   \\
\>\rm{05}\>\>\>Problem ``make/function''   \\
\>\rm{06}\>\>\>\>solve\_univariate $\;\left(\left(\frac{u}{2}\right)^2+\left(\frac{v}{2}\right)^2=r^2\;v\right)$
                                  \`\it{$<$collapsed$>$} \\
\>\rm{07}\>\>\>\>$L = \left[v=2\cdot\sqrt{r^2-\left(\frac{u}{2}\right)^2}\;v=-2\cdot\sqrt{r^2-\left(\frac{u}{2}\right)^2}\right]$\\
\>\rm{08}\>\>\>\>                 \`$v \in \{0<..<r\}$ \\
\>\rm{09}\>\>\>\>$A = 2\cdot u\cdot v-u^2$\\
\>\rm{10}\>\>\>\>                 \`\it{Substitute} $\;v$ \\
\>\rm{11}\>\>\>$A\;u = 2\cdot u\cdot\sqrt{r^2-\left(\frac{u}{2}\right)^2} - u^2$\\
\>\rm{12}\>\>\>differentiate $\left(A\;u = 2\cdot u\cdot\sqrt{r^2-\left(\frac{u}{2}\right)^2} - u^2\;u\right)$\\
\>\rm{13}\>\>\>\>$A^\prime\;u = \frac{d}{du}\left( 2\cdot u\;\sqrt{r^2-\left(\frac{u}{2}\right)^2} - u^2 \right)$\\
\>\rm{14}\>\>\>                   \`$\frac{d}{du} (u_1-u_2) = \frac{d}{du} u_1 - \frac{d}{du} u_2$\\
\>\rm{15}\>\>\>\>$A^\prime\;u = \frac{d}{du}\left( 2\cdot u\;\sqrt{r^2-\left(\frac{u}{2}\right)^2} \right) - \frac{d}{du}u^2$\\
\>       \>\>\vdots \\
\>\rm{29}\>$[u=6.75 v=12.25]$
\end{tabbing}}

\noindent
The ``Solution'' iterates through (a) interactively constructing steps 
towards a solution of the problem (``Solution'' phase) and through (b) specifying 
sub-``Problems'' (``Specification'' phase); probably for more than one sub-problem.
The first encounter is Problem ``make/function'' in line \rm {05} immediately followed by
another problem ``solve\_univariate'' in line \rm{06}, which appears in
a format like in algebra systems (and where the dialogue-module allows
to skip long-winded ``Specification'').

At the right margin there are (beyond $<$\textit{collapsed}$>$) the 
justifications for the steps constructing solutions; in line \rm{09} the
negative element is dropped from the set $L$ of solutions because it is
not element of the interval $\{0<..<r\}$ known from formalisation $F_I$;
in line~\rm{10} the justification is
the primitive operation ``Substitute $v$'', which in line \rm{11}
constructs the solution for Problem ``make/function''.

In line \rm{12} ``differentiate'' is again a function known from 
computer algebra (and thus the respective ``Specification'' is skipped).
The function creates a sequence of steps in a medium number;
here the justification is given by the well-known theorem 
$\frac{d}{du} (u_1-u_2) = \frac{d}{du} u_1 - \frac{d}{du} u_2$. 
Finally at a certain point
an engineer will be interested in concrete numbers for $u$ and $v$ 
and will decide to switch to floating-point numbers in some 
interactive ``Specification''.

\paragraph{Extent of realisation and accessibility:} The dialogue-module presently is a stub 
necessary to drive field studies. The dialogue will be guided by a user-model, 
which will call for expertise in learning theory and requires a larger user group for
iterative development, driven by systematic feedback from users.
Accessibility design will be concerned with tools for switching between detail
and survey, in particular in complex problem solutions with several 
sub-problems.

\section{Software Components for Future Development}\label{techno}
The fact that VSCode is accessible does not mean, that accessible Isabelle/VSCode
is already done. Also the present front-end jEdit required work of more than a
decade to make Isabelle/jEdit a useful tool as is.

\subsection{Status-quo of Interactive Prover Technology}\label{state-prover}

We intend to continue long-standing traditions of interactive proof
assistants, which started in the 1970s as rather modest
Read-Eval-Print Loop (REPL) of the LCF proof checker
\cite{LCF-to-Isabelle-HOL:2019}, and are manifest today in a handful
of big and successful systems (with associated communities): HOL4,
Coq, Isabelle, Lean Prover etc.  Our platform of choice is Isabelle
--- compared to the other proof assistants, it is notable for two main
reasons:

\begin{compactenum}

\item \textbf{Isabelle/AFP} --- The Archive of Formal Proofs. This is
  a collection of formal theory developments in Isabelle, presented as
  an online Journal (\url{https://isa-afp.org}). All entries are
  formally checked against a recent version of the Isabelle proof
  assistant. Users may \emph{explore} mathematical content via regular
  web-browser technology, or download entries to \emph{edit} in the
  Isabelle development environment.

\item \textbf{Isabelle/PIDE} --- the Isabelle Prover IDE framework,
  based on Isabelle/Scala (on the Java platform). Conceptually, it is
  a semantically enhanced source text editor with purely functional
  back-end (Isabelle/ML) and various text editor
  front-ends. Isabelle/jEdit is the best-known and best-developed
  front-end. Isabelle/VSCode is an alternative, but still lagging
  behind Isabelle/jEdit significantly.
  
\end{compactenum}

The Isabelle eco-system provides many more technologies to support
mathematical proofs, including automated provers (ATPs and SMTs). All
of this is well-integrated for the standard operating systems (Linux,
Windows, macOS): applications can benefit from the Isabelle system
distribution, and include themselves easily as add-on components.

Moreover, Isabelle tool developers can use existing IDE technology
like Isabelle/jEdit for mathematical programming in Isabelle/ML, and
IntelliJ IDEA for systems programming in Isabelle/Scala
\cite[p.~60]{isa-system-2021}.

\subsection{Prospective Technology for Accessible Mathematics}\label{prospect-tech}

For our end users, we need to focus on software components that support
accessibility properly; various technical issues have been discussed
in \S\ref{state}. Subsequently, we sketch an architecture specifically
for the Isabelle framework and its {\isac} application, together with
open questions of research and development.

The key idea is to re-use HTML/CSS/JavaScript web technology, as implemented
in the particular Chrome browser engine (by Google) that is bundled
with the VSCode editor (by Microsoft). This results in a conventional
desktop application, (\emph{not} web application), which avoids the
typical HTML incompatibilities known as ``browser hell''.

Specifically, we shall extend and improve two software projects that
are based on the Isabelle platform: Isabelle/VSCode and \sisac.

\begin{compactitem}

\item Isabelle/VSCode is an extension for the VSCode text
  editor. It combines the Microsoft Language Server Protocol
  (LSP)\footnote{\url{https://microsoft.github.io/language-server-protocol/}}
  with the Isabelle Prover IDE framework (PIDE)
  \cite{Wenzel:2019:MKM}.  Since VSCode allows to integrate free-form
  HTML views, our add-ons to Isabelle/VSCode will be twofold:

  \begin{compactenum}

  \item Editor support, using specific VSCode extension interfaces.

  \item Browser support, using generic HTML/CSS/JavaScript technology.
    
  \end{compactenum}

\item The \sisac{}
  project\footnote{\url{https://hg.risc.uni-linz.ac.at/wneuper/isa}}
  implements a concept of ``Next-Step Guidance''~\cite{EPTCS-wn-20} of
  mathematical reasoning, based on Isabelle theory
  content~\cite{wn:proto-sys}. It was historically built on the
  Isabelle command-line with add-on Java GUI (now discontinued). The
  plan is to retrofit \sisac{} into the interactive document-model of
  Isabelle/PIDE/VSCode, and thus let it participate in its support for
  accessibility.
  
\end{compactitem}

\subsection{Open Questions}\label{open-technologies}

Some of the questions shall be answered during master theses gaining experience 
for the detailed project plan,
some during writing the project proposal and finally some during the
project itself.

\begin{compactenum} 
\item\textbf{Workflow} in investigating, coding and testing between 
Isabelle/VSCode (production $+$ development in TypeScript), IntelliJ IDEA 
(development in Scala) and Isabelle/jEdit (production $+$ development in 
Isabelle/ML) --- How can we make such work-flows efficient?
\item\textbf{Adding an ``Outline view''} in Isabelle/VSCode according to UR.3 
--- does this involve extending the TypeScript plugin / LSP and JSON / 
Isabelle/Scala / Isabelle/ML?
\item\textbf{Mapping Braille standards} with ASCII string representation of 
proper (sub-) terms according to UR.1 --- which changes does this require in the 
TypeScript plugin / in LSP and JSON / in Isabelle/Scala / in Isabelle/ML?
\item\textbf{Improving Isabelle/VSCode} --- which improvements would serve 
sighted and blind users most with least effort?
\item\textbf{Isabelle's databases} for theories, theorems, constants, 
PIDE markup, etc. --- can these be used for surveys according to UR.2? Can 
certain static items like explanations be added to these databases?
\end{compactenum}
The questions indicate demand of expertise in development of accessibility, of
technologies like JSON, LSP, etc and of certain Isabelle technologies. According 
to the above open questions, efforts in development will be distributed over 
respective expertise.

\section{A Vision for Educational Math Software}\label{edu-goal}
Since on the one hand \S\ref{access-math} stated ambitious user requirements with respect to
accessibility, \S\ref{edu-soft-math} complained about lack of certain software
support and on the other hand in \S\ref{case-study} details have been given 
of how gaps will be filled and
\S\ref{techno} introduced software components rarely used
for development of software of the envisaged kind, this section exposes
visions, which are driving the MAWEN project.

\medskip
The traditional way of using paper and pencil to write down mathematical 
contents, and much more to do mathematical calculations, is still the predominant 
method among all people who deal with mathematics in education. This is especially 
remarkable if we consider the tremendous changes modern information technology 
brought into society over the past few decades. Although computer algebra 
systems such as Mathematica are widely used among mathematicians, these systems 
cannot help a person learning to do a calculation by hand, since they will do the 
calculation for him/her. A system where the computer with its ever increasing 
possibilities to support the human brain would help a person to actually do a 
calculation is still missing, not only, as outlined here, for the group of blind 
or visually impaired people, but for the sighted mainstream as well --- one will 
publish, or distribute, mathematical content in a digital format, but, as soon 
as one calculates by him/herself, it is generally paper and pencil which is used.

\par Of course the situation is changing step by step and faster due to the 
growing number of touch/pen based systems including OCR for math, digital 
math-ink, which will allow developing a new digital culture of interactively 
doing math including and integrating the interaction with tools as computer 
algebra, theorem provers and tutoring systems. These new and innovative systems, 
inviting to switch from paper/pencil to digital doing math, provide a 
much better and much richer source for supporting navigation and in particular 
doing math for blind and low vision people: What was implicit to the visual 
procedures and workflows of doing math with pencil and paper, now has to be 
formal and therefore explicitly described in a machine readable and 
processible manner. This forms a rich digital base, to which accessibility can 
hopefully be efficiently added and which allows to implement assisting and supportive 
functionalities better in accordance with the needs of blind and low vision 
people. 

\medskip
\par In spite of the undisputed advantages of the traditional paper-and-pencil 
method, it is our conviction that the computer has a strong potential to help in 
a calculation, be the user sighted, visually impaired, or blind. We thus 
envision a scenario, where technical support apparently needed by the relatively 
small group of blind people may as well be of great use and value for all people 
who are dealing with mathematics. We see this formal mark-up of data and of the procedures 
of doing math as a first step towards a much more personalised approach to doing 
math problem solving, which is able to much better respect the broad diversity of 
users in terms of physical, sensory, psycho-cognitive and social factors.

\section{Summary and Conclusions}\label{conclusion}
This paper, describing preparations for a concrete project (named MAWEN), 
restricts the scope of accessibility to serve blind and visually impaired students
without further impairments. 

The projects strengths are provided by two novelties: (I) a larger coverage of 
software support for learning mathematics implemented by the \sisac-prototype
and (II) the technical features of Isabelle/VSCode supporting accessibility.
In order to make the strengths of the novelties comprehensible, a survey on 
preparatory work in MAWEN and a 
survey of demand in educational practice is given in \S\ref{access-math}.
A survey of demand for mathematics education in general is given in 
\S\ref{edu-soft-math}, which identifies a lack of coherent software support in a 
central point of math education, in step-wise solving mathematical problems 
(of "real world" problems at high school as well as of real engineering problems 
in academic courses).
\S\ref{proof} identifies more need in mechanical proof with ongoing advancement
of Formal Methods in industry --- again pointing out lack of software support
for early experiences with prerequisites of mathematical proof, for instance
matching.

\emph{Since effectiveness of software support depends on the extent of opportunities
for computer usage, extending the coverage of computer support is essential ---
MAWEN plans to widely fill some gaps and diminish the lack of appropriate
software for mathematics education, for blind students in inclusive learning
scenarios.} 

\emph{Development of such widely usable software is possible with reasonable effort, 
because it can start from results of extensive prototyping in the \sisac{} project
and because of the structural affinity of Braille display and of Isabelle's 
representation of formulas --- linear structure in both cases (for instance 
$(x+1)/(y-2))$ instead of $\frac{x+1}{y-2}$) and accessibility inherited from 
VSCode.}

Accomplishment of the respective user requirement UR.1 in \S\ref{user-requ}
by the front-end technology in Isabelle/VSCode is discussed in \S\ref{techno}
together with other technical features of Isabelle, advantageous for 
implementation of widely usable software for inclusive mathematics education.
Respective realisability is underpinned by an extensive case study
in \S\ref{example}..\S\ref{iterating} and \S\ref{edu-goal} summarises the
vision guiding the MAWEN project.

The authors hope that the description of the preparations convince the kind reader
of the realisability of our high expectations, that they appeal to European
partners for collaboration in joint projects, and last not least that they
convince funding agencies that extending the coverage in software support of blind and
visually impaired people is worth some resources.

\bibliographystyle{eptcs}
\bibliography{bibliography}

\end{document}